\documentclass[%
reprint,
superscriptaddress,
showpacs,preprintnumbers,
 amsmath,amssymb,
 aps,
]{revtex4-1}

\usepackage{graphicx}
\usepackage{dcolumn}
\usepackage{bm}

\usepackage{epsfig}
\usepackage{hyperref}
\begin{document}

	\title{Field-tuned spin excitation spectrum of $k\pi$ skyrmion}
	
	\author{Chengkun Song}%
	\affiliation{
		Key Laboratory for Magnetism and Magnetic Materials of the Ministry of Education, Lanzhou University, Lanzhou, 730000, People’s Republic of China
	}%
	\author{Yunxu Ma}%
	\affiliation{
		Key Laboratory for Magnetism and Magnetic Materials of the Ministry of Education, Lanzhou University, Lanzhou, 730000, People’s Republic of China
	}%
	\author{Chendong Jin}%
	\affiliation{
		Key Laboratory for Magnetism and Magnetic Materials of the Ministry of Education, Lanzhou University, Lanzhou, 730000, People’s Republic of China
	}%
	\author{Haiyan Xia}%
	\affiliation{
		Key Laboratory for Magnetism and Magnetic Materials of the Ministry of Education, Lanzhou University, Lanzhou, 730000, People’s Republic of China
	}%
	\author{Jinshuai Wang}%
	\affiliation{
		Key Laboratory for Magnetism and Magnetic Materials of the Ministry of Education, Lanzhou University, Lanzhou, 730000, People’s Republic of China
	}%
	
	\author{Jianbo Wang}%
	\affiliation{
		Key Laboratory for Magnetism and Magnetic Materials of the Ministry of Education, Lanzhou University, Lanzhou, 730000, People’s Republic of China
	}
	\affiliation{
		Key Laboratory for Special Function Materials and Structural Design of the of Ministry of Education, Lanzhou University, Lanzhou, 730000, People's Republic of China
	}%
	\author{Qingfang Liu}
	\email{liuqf@lzu.edu.edu}
	\affiliation{
		Key Laboratory for Magnetism and Magnetic Materials of the Ministry of Education, Lanzhou University, Lanzhou, 730000, People’s Republic of China
	}%
	
	
\date{\today}
	
\begin{abstract}
We study spin wave excitation modes of $k\pi$ skyrmion in a magnetic nanodot under an external magnetic field along $z$-direction using micromagnetic simulations based on Landau-Lifshitz-Gilbert equation. We find that a transition of $k\pi$ skyrmion to other skyrmion-like structures appears under some critical external field, the corresponding spin wave excitations are simulated for each state in the process of applying magnetic field. For skyrmion, the frequencies of excitation modes increases and then decreases with the low frequency modes splitting at a critical magnetic field. In addition to the well known two in-plane rotational modes and a out-of-plane breathing mode of skyrmion, a higher number of excitation modes are found with increasing $k$ ($k=2, 3$). The excitation modes vary as a function of magnetic field, and the excitation frequencies for different modes exhibit a rapid or slight change depending on the field induced change of magnetization profile. Our study indicates the rich spin wave excitations for $k\pi$ skyrmion and opens a possibility in theoretical or experimental investigation of magnonics application.
\end{abstract}

\maketitle


\section{Introduction}
As a topological protected local whirl of spin configurations, magnetic skyrmion attracts a lot of attention since it has been observed in MnSi~\cite{muhlbauer2009skyrmion,yu2015variation,pappas2009chiral}. The direction of spins inside the skyrmion points up (down), while the surrounding spins of the ferromagnetic background are pointing in opposite direction which is down (up)~\cite{nagaosa2013topological,sampaio2013nucleation,fert2013skyrmions}. Depending on the chirality of the domain wall determined by the Dzyaloshinsky-Moriya interaction (DMI) type~\cite{dzyaloshinsky1958thermodynamic,moriya1960anisotropic}, which separates the spins inside of the skyrmion and the surrounding spins, there are two distinct types skyrmions with rotation symmetry defined as N$\mathrm{\acute{e}}$el skyrmion and Bloch skyrmion~\cite{kezsmarki2015neel,tomasello2014strategy}. They have been observed in chiral bulk helimagnets in the presence of bulk DMI~\cite{bogdanov1989thermodynamically,rossler2011chiral,tonomura2012real,tokunaga2015new} and in ultrathin magnetic films contacted with a heavy metal layer in the presence of interfacial DMI~\cite{leonov2016properties,rohart2013skyrmion,jiang2015blowing,jiang2016mobile}. Recently, antikyrmions with rotation symmetry breaking have been demonstrated in magnetic materials belonging to crystallographic classes $D_{2d}$ and $S_4$~\cite{koshibae2016theory,nayak2017magnetic,hoffmann2017antiskyrmions,song2018magnetic}. For skyrmions or antiskyrmion, the angle of progressive magnetization rotation is 1$\pi$, where the angle is defined as the number of sign changes of the perpendicular magnetization when moving along the radial direction.

Skyrmions are promised in rich application, such as racetrack memory~\cite{jiang2015blowing,woo2016observation,sampaio2013nucleation,zhang2016magnetic,song2017skyrmion}, spin transfer nano-oscillator~\cite{garcia2016skyrmion,zhang2015current}, transistor~\cite{zhang2015magnetic,fook2016gateable} or in other spintronic devices~\cite{zhang2015magnetic,prychynenko2018magnetic,everschor2018perspective}. In addition to the application on spintronics, a possible application for skyrmions is in the field of magnonics. It is important in fundamental physics and manipulation to investigate the spin wave modes of skyrmions~\cite{mochizuki2012spin,onose2012observation,mruczkiewicz2017spin,mruczkiewicz2018azimuthal}. The spin wave modes of skyrmions were investigated in skyrmion lattice and isolated skyrmion in confined magnetic infrastructures~\cite{kim2014breathing,mruczkiewicz2018azimuthal}. Typical spin wave excitation modes include a couple of rotation modes for in-plane microwave magnetic field and a breathing mode for out-of-plane microwave magnetic field. 

Other than the skyrmions or antiskyrmions with 1$\pi$, some skyrmion-like magnetization structures with rotation angle $k\pi$ are experimentally and theoretically investigated in recent years, where the DMI is responsible for the stabilization~\cite{rozsa2018localized,mulkers2016cycloidal,beg2015ground,finazzi2013laser,finizio2018thick,streubel2015magnetization}. The perpendicular magnetization of $k\pi$ skyrmion rotates multiple times along the radial direction with $k > 1$ respect to skyrmion with 1$\pi$. Compared to the rich investigations of skyrmion, skyrmion-like structures with higher $k$ attract much less attention. The generation and the dynamics of 2$\pi$ skyrmion under current are studied by Liu et al~\cite{liu2015static} and Zhang et al~\cite{zhang2016control}. Moreover, the 2$\pi$ skyrmion can be driven to propagate under a spin wave~\cite{shen2018motion,li2018dynamics}. Recently, the controlled creation of $k\pi$ skyrmions on a discrete lattice are investigated in Ref.~\cite{hagemeister2018controlled}, while the localized spin wave modes under an external magnetic field are still unexplored.

Here, we investigate the influence of external perpendicular magnetic field on spin wave excitation modes of $k\pi$ skyrmion in a circular magnetic nanodot. The parameters of the system is taken to ensure that the $k\pi$ skyrmion ($k=1, 2, 3$) can exist as ground states separately. By applying magnetic field, the transformations between skyrmions with different $k\pi$ are studied. We also demonstrated that the spin wave excitation modes are quite different for $k\pi$ skyrmion, which are more complicated than that of skyrmion. Using a microwave magnetic field, different excitation modes are depicted. When $k=2, 3$, some mixed modes appear except for the counterclockwise (CCW) and clockwise (CW) excitation modes for an isolated skyrmion in a nanodot, the frequency of excitation modes as a function of external field are depicted as well. These results enable us to distinguish and classify skyrmions with different $k\pi$, and maybe promise in the application of magnonics by controlling the magnetization dynamics using spin wave.

\section{Simulation model}

The simulation system considered in our study is a circular ferromagnetic nanodot, where the radius and thickness are fixed as $R = 60 \ \mathrm{nm}$ and $t = 0.6 \ \mathrm{nm}$. We use Mumax3 code to perform micromagnetic simulation~\cite{Vansteenkiste2014}. The dynamics of $k\pi$ skyrmion and its spin excitation spectrum are governed by Landau-Lifshitz-Gilbert (LLG) equation of magnetization~\cite{fert2013skyrmions,sampaio2013nucleation}
\begin{equation}
\frac{\partial \mathbf{m}}{\partial t}=\gamma \mathbf{m}\times\mathbf{H}_{\mathrm{eff}} + \alpha \mathbf{m}\times \frac{\partial \mathbf{m}}{\partial t}
\end{equation}
where $\mathbf{m}$ is unit magnetization vector, which is defined as ${\mathbf{M}}/{M_\mathrm{s}}$ and $M_\mathrm{s}$ is the saturation magnetization. $\gamma$ is the gyromagnetic ratio, $\alpha$ is the Gilbert damping parameter. $\mathbf{H}_{\mathrm{eff}}=\mathbf{H}_{\mathrm{exch}}+\mathbf{H}_{\mathrm{anis}}+\mathbf{H}_{\mathrm{DMI}}+\mathbf{H}_{\mathrm{ext}}+\mathbf{H}_{\mathrm{d}}$ is the effective magnetic field of the system, which includes the Heisenberg exchange field $\mathbf{H}_{\mathrm{exch}}$, the uniaxial perpendicular magnetic anisotropy field $\mathbf{H}_{\mathrm{anis}}$, the Dzyaloshinskii-Moriya exchange field $\mathbf{H}_{\mathrm{DMI}}$, the external magnetic field $\mathbf{H}_{\mathrm{ext}}$ and the demagnetizing field $\mathbf{H}_{\mathrm{d}}$. 

The simulation code resolves the LLG equation using finite difference method and the unit cell size is set as 1 nm $\times$ 1 nm $\times$ 0.6 nm. To obtain the initial equilibrium $k\pi$ skyrmion state ($k=1,2,3$), we use the same magnetic parameters for Pt/Co multilayer, which has been found to exhibit magnetic skyrmions~\cite{moreau2016additive,zhang2018direct}. The micromagnetic simulation parameters are chosen as~\cite{sampaio2013nucleation}: exchange stiffness $A = 15\times10^{-12} \ \mathrm{J/m}$, uniaxial magnetocrystalline anisotropy $K_u = 0.8\times 10^6 \ \mathrm{J/m^3}$, saturation magnetization $M_\mathrm{s}=580\times10^3 \ \mathrm{A/m}$. The interfacial DMI is considered and the DMI constant is set as $D=4.0 \ \mathrm{mJ/m^2}$, in which the energy density as a sum of Lifshitz invariant is~\cite{sampaio2013nucleation,ezawa2011compact}
\begin{equation}
\xi = D(\mathbf{m}\cdot \nabla m_z-m_z\nabla\cdot \mathbf{m}).
\end{equation}

\begin{figure}[!htb]
	\begin{center}
		\epsfig{file=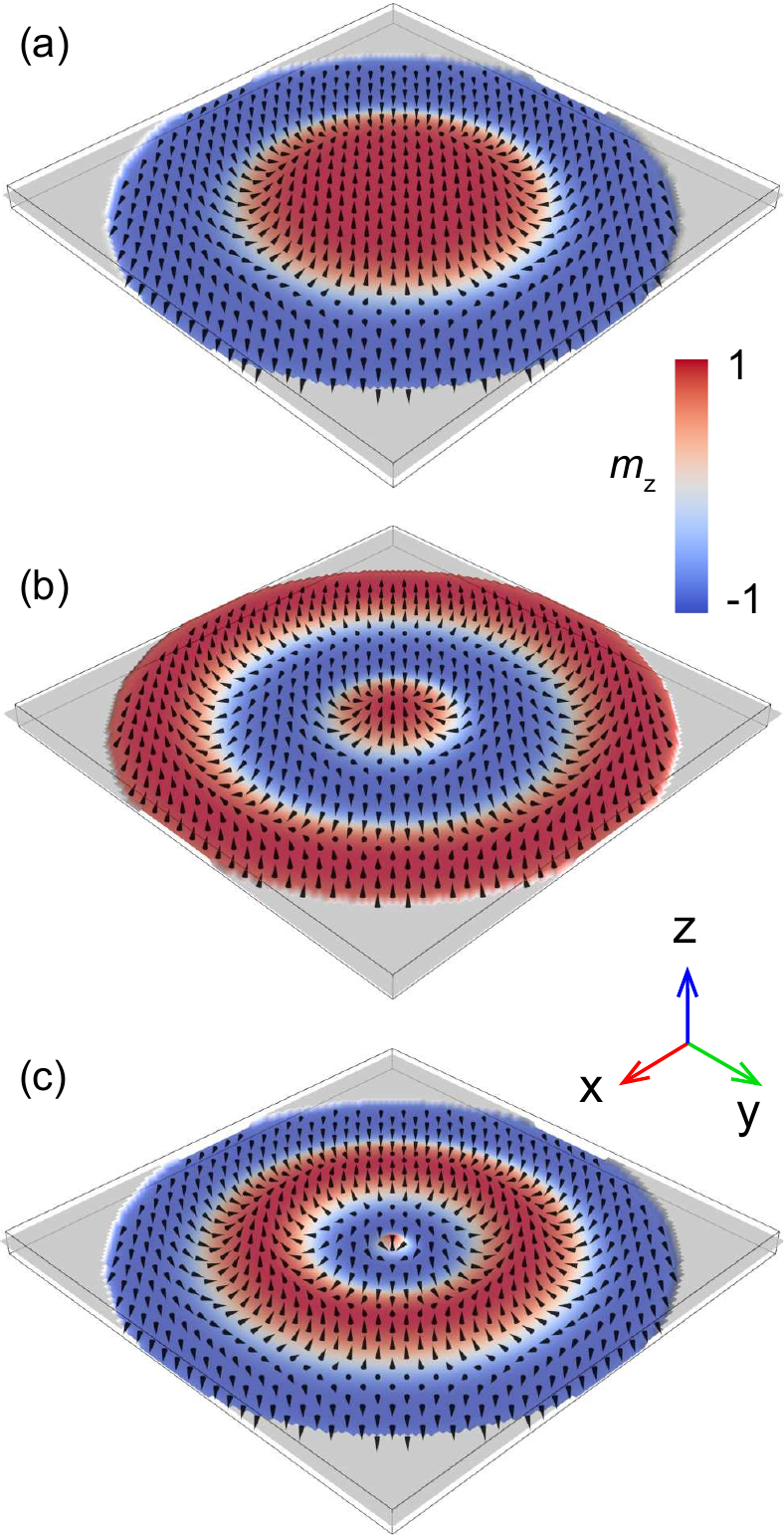,width=8 cm} \caption{Three different magnetization configurations of the circular magnetic nanodot, (a) skyrmion, (b) $2\pi$ skyrmion and (c) $3\pi$ skyrmion. The magnetization components in the center of the nanodot are all along $z$ direction.}
		\label{fig1}
	\end{center}
\end{figure}

Three different $k\pi$ skyrmion states ($k = 1, 2, 3$) in the nanodot are relaxed as the initial states, respectively. The parameters taken in our system ensure the skyrmion, $2\pi$ skyrmion and $3\pi$ skyrmion can exist as the ground states. As shown in Fig.~\ref{fig1}, the magnetization profiles of three skyrmion-like topological protected magnetization states are depicted with the rotation angle (a) $1\pi$ (b) $2\pi$ and (c) $3\pi$. Color map characterizes the direction of out-of-plane magnetizations $\mathbf{m}_z$, where the red color represents the direction along $z$ direction and the blue color represents the direction along -$z$ direction. For compassion of $k\pi$ skyrmions, we set the magnetization in the center along $z$ direction.

In the following simulation, the external perpendicular magnetic field is applied along $z$-direction for the above three magnetization configurations, which varies from 0 mT to 300 mT. In order to investigate the spin excitation dynamics of $k\pi$ skyrmion, we consider another uniform magnetic field pulse along $x$ or $z$ direction as excitation field depending on the spin excitation modes with different symmetry we are interested. The excitation field exhibits low amplitude and have a time dependence with $sinc$ function type, $H_{x(z)} = H_0sinc(2\pi ft) = sin(2\pi ft)/(2\pi ft)$, where the amplitude is represented by $H_0$ which is 10 mT and the cut-off frequency $f$ is set as 50 GHz. After applying the excitation magnetic field, the magnetization components as a function time and space are transformed as a function of frequency using Fourier transform, thus we can acquire the corresponding frequencies of different spin excitation modes, as well as the spatial distribution of the Fast Fourier Transformation (FFT) power at specific oscillation frequency of magnetization components. In the following simulation, the damping constant $\alpha$ is 0.01 unless othrerwise specified.

\section{Results and discussion}

\subsection{Field dependent $k\pi$ skyrmion size}

\begin{figure}[!htb]
	\begin{center}
		\epsfig{file=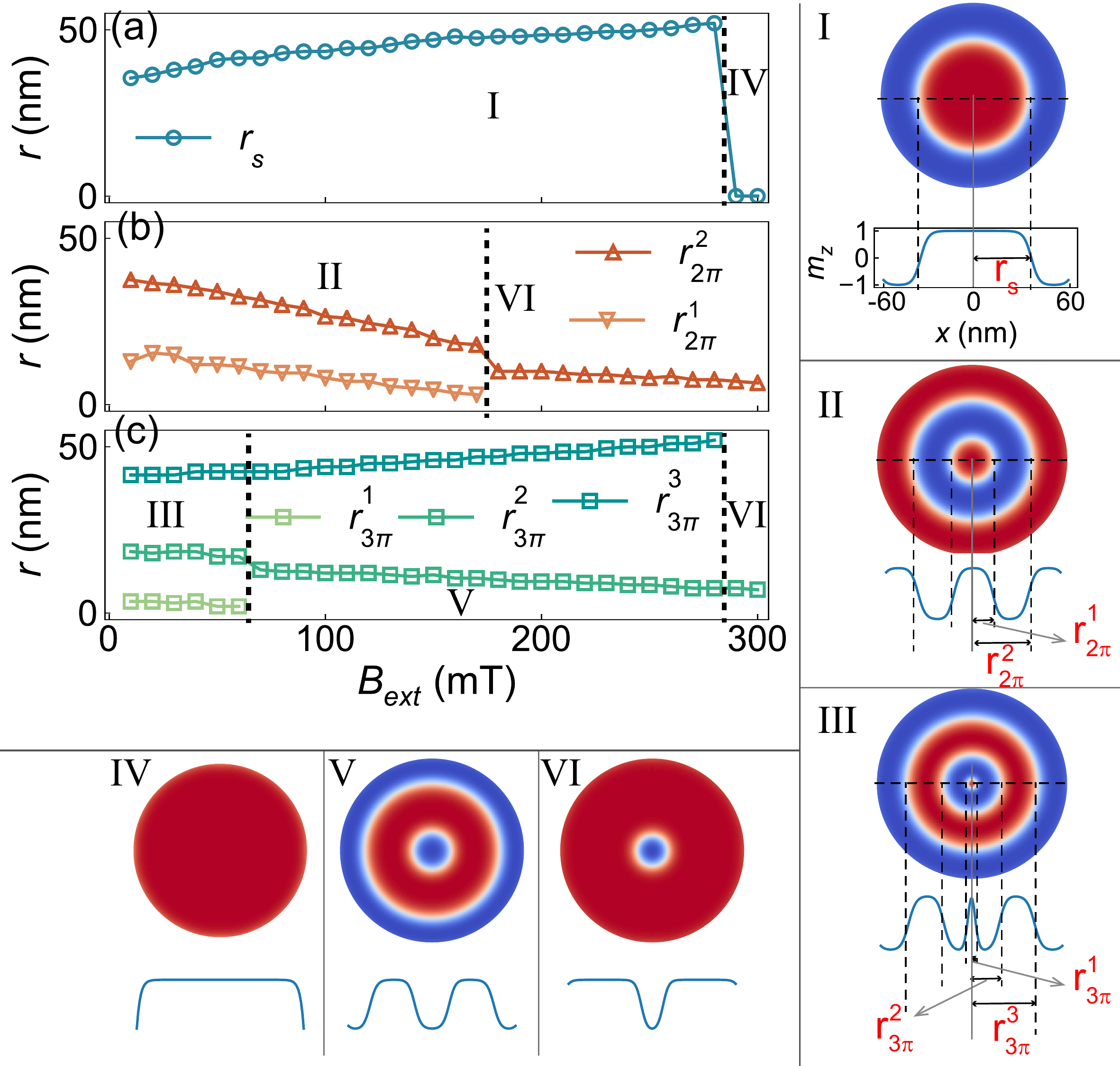,width=8 cm} \caption{The radius of $k\pi$ skyrmion as a function of magnetic field normal to the nanodot plane for (a) skyrmion, (b) $2\pi$ skyrmion and (c) $3\pi$ skyrmion. The magentization profiles in different magnetic field range are marked as I, II, III, IV, V and VI. I, II and III correspond to skyrmion, $2\pi$ skyrmion and $3\pi$ skyrmion we are interested. The corresponding magnetization profiles of $\mathbf{m}_z$ are depicted under the magnetization configurations of different states, as well as the radius determined.}
		\label{fig2}
	\end{center}
\end{figure}

Firstly, we discuss the size of $k\pi$ skyrmion as a function of external magnetic field as well as the transitions between different magnetization states, where the magnetic field is applied normal to the nanodot along $z$ direction. As shown in Fig.~\ref{fig2}, the initialized magnetization profiles of skyrmion, $2\pi$ skyrmion and $3\pi$ skyrmion are marked with capitalized Roman numerals I, II, III, respectively. The magnetization states represented by IV, V and VI are the transformed states for $k\pi$ skyrmion at the critical magnetic field. Color map plots of the magnetization states are same as that shown in Fig.~\ref{fig1}. The radius of skyrmion is determined by the radius of the rings in which composed by a domain wall (N$\mathrm{\acute{e}}$el wall), which is the transition area between two out-of-plane magnetization regions and represented by $r_s$. For $k\pi$ skyrmion, the numbers of the rings equal to $k$. Similar to the definition for skyrmion, the radius of $2\pi$ skyrmion is characterized by two variables $r_{2\pi}^1$ and $r_{2\pi}^2$, which are defined from inner to outer part. Moreover, three variables $r_{3\pi}^1$, $r_{3\pi}^2$ and $r_{3\pi}^3$ represents the radius of $3\pi$ skyrmion named from the inner ring to the outer ring, as shown in the right part of Fig.~\ref{fig2}.

The radius of $k\pi$ skyrmions are shown in Fig~\ref{fig2} for (a) skyrmion, (b) $2\pi$ skyrmion and (c) $3\pi$ skyrmion as a function of magnetic field. The direction of the magnetic field is parallel to the direction of magnetization in the center of the nanodot, which is in the range of 0 mT to 300 mT. For skyrmion, the radius $r_s$ increases rapidly with increasing the magnetic field under 160 mT, and then it increases slowly until 280 mT. This is due to the competition of the field-induced skyrmion expanding and boundary induced reduction, the skyrmion keeps a weak expansion with increasing field. When the field increases larger than 280 mT, the magnetization state in the nanodot collapse to a uniform state (IV) with magnetization components along $z$ direction. While for $2\pi$ skyrmion, $r_{2\pi}^1$ and $r_{2\pi}^1$ decreases as a function of magnetic field. Compared to the energy of the inner ring, the energy of the outer ring is more higher. It is noted that the $2\pi$ skyrmion collapses to an another skyrmion (VI) when $B_{ext} = 180$ mT, in which the polarity is opposite to the skyrmion marked as I. For $3\pi$ skyrmion, $r_{3\pi}^1$ and $r_{3\pi}^2$ decrease with increasing field until 60 mT, while $r_{3\pi}^3$ increases. When the magnetic field is larger than 60 mT, $3\pi$ skyrmion transforms to a $2\pi$ skyrmion state (V) with the inner ring vanishing, for which the polarity is opposite to the $2\pi$ skyrmion state II. The radius corresponding to the second ring ($r_{3\pi}^2$) and the outer ring ($r_{3\pi}^3$) decreases and increases as a function of field with increasing to $B_{ext} = 280$ mT, respectively. Then, it collapses to a skyrmion sate (VI) when the field exceeds 280 mT. 

\subsection{Field dependent spin excitation of $k\pi$ skyrmion}

\begin{figure}[!htb]
	\begin{center}
		\epsfig{file=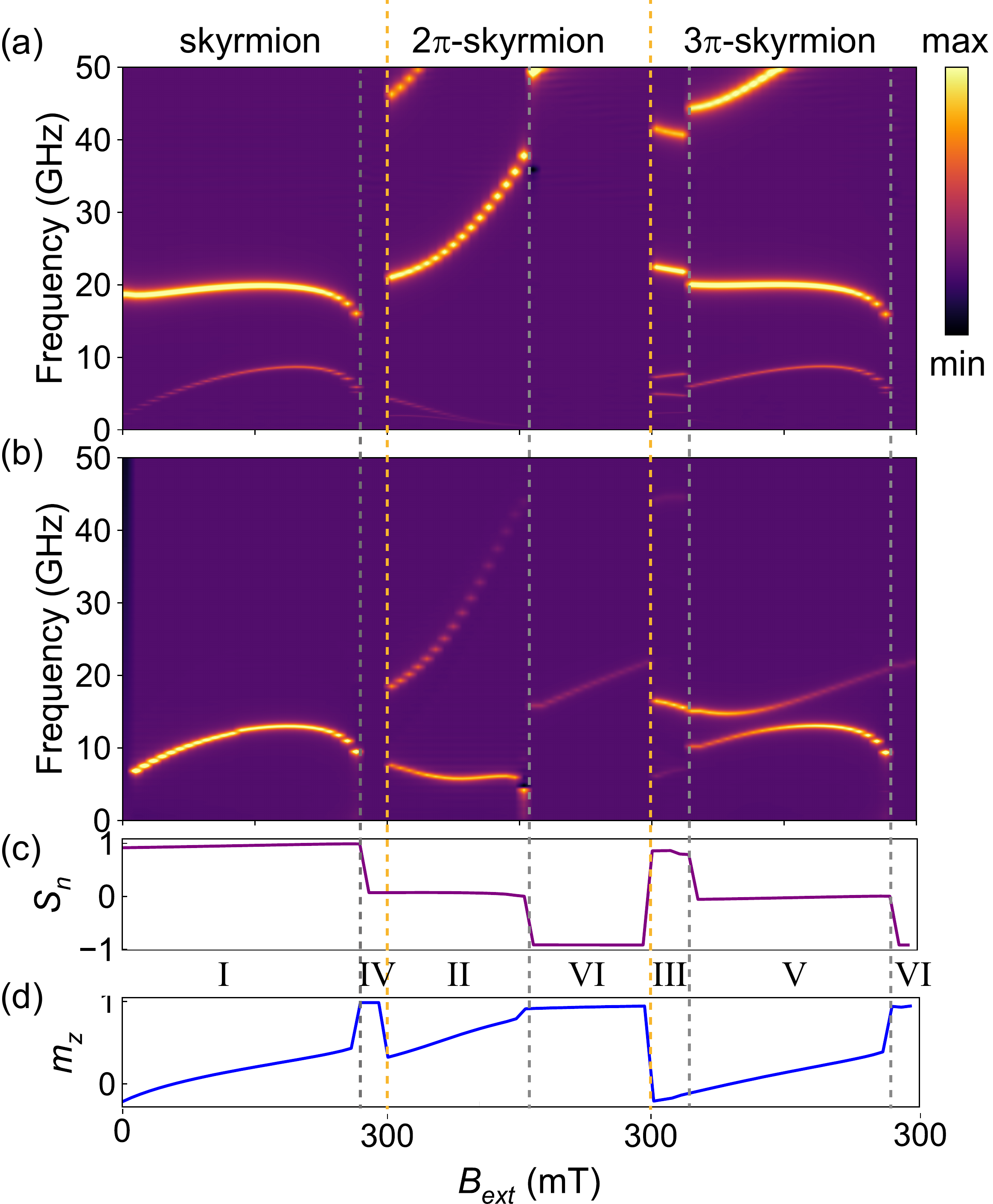,width=8 cm} \caption{Frequency of skyrmion, $2\pi$ skyrmion and $3\pi$ skyrmion as a function of magnetic field in the range of 0 mT to 300 mT, which obtained through FFT of corresponding magnetization components by applying uniform magnetic field pulse with $sinc$ type in (a) in-plane direction and (b) out-of-plane direction. They are separated by yellow dash line, and the gray dashed lines represented the critical magnetic field where the magnetization state collapses to another state. The static properties of different magnetization states are characterized by (c) skyrmion number $S_n$ and (d) magnetization component along $z$ direction $\mathbf{m}_z$.}
		\label{fig3}
	\end{center}
\end{figure}
After investigating $k\pi$ skyrmion size as a function of magnetic field, in this section, we focus on the spin eigenmodes for $k\pi$ skyrmion as well as the transition of different magnetization states under perpendicular magnetic field. The spin excitation modes include the eigenmode with radial symmetry, also known as breathing mode, and the eigenmode with broken radial symmetry which known as azimuthal modes containing CW and CCW directions. CCW and CW modes are a precession of the topological center around the ground state in the nanodot center. To excite the breathing mode exhibits radial symmetry, an out-of-plane magnetic field pulse was applied. The azimuthal modes include gyrotropy in CW and CCW directions can be excited by an in-plane magnetic field pulse. The calculated frequency of spin excitation modes as a function of magnetic field for skyrmion, $2\pi$ skyrmion and $3\pi$ skyrmion are presented in Fig.~\ref{fig3}. Fig.~\ref{fig3} (a) shows the azimuthal modes for in-plane excitation magnetic field and Fig.~\ref{fig3} (b) depicts the breathing mode for out-of-plane excitation magnetic field. The regions with different $k\pi$ skyrmion are separated with perpendicular yellow dashed lines. In each region, the magnetic field increases from 0 mT to 300 mT with a step of 10 mT. As we have mentioned, $k\pi$ skyrmion states collapse to another magnetization configurations at some critical magnetic fields. These transformed states can be differentiated using topological number $S_n$ and magnetization components along $z$ direction $\mathbf{m}_z$, as shown in Fig.~\ref{fig3} (c) and (d), respectively. The capitalized Roman numerals I-VI represent static  magnetization states formed in specific magnetic field, which have been depicted in Fig.~\ref{fig2}. The simulated static magnetization configurations in the nanodot are characterized by the topological number, named skyrmion number~\cite{nagaosa2013topological}
\begin{equation}
S_n = \frac{1}{4\pi}\int \int \mathbf{m}\cdot(\frac{\partial \mathbf{m}}{\partial x}\times\frac{\partial \mathbf{m}}{\partial y})dxdy.
\end{equation}

In ideal situation, the skyrmion number is 1 for skyrmion, 0 for $2\pi$ skyrmion and 1 for $3\pi$ skyrmion in our simulation system. However, it is not exactly equal to them due to the DMI induced magnetization rotation at the boundary of the restricted nanodot. It should be mentioned that the skyrmion number is equal for state II and state V, which is 0, while the magnetization states are different. They are all called $2\pi$ skyrmion with opposite magnetization component in the center of the nanodot. Thus, we use another static property that is average magnetization component along $z$ direction $\mathbf{m}_z$ to characterize different magnetization configurations, as depicted in Fig.~\ref{fig3} (d). We can see that $\mathbf{m}_z > 0.5$ for $2\pi$ skyrmion (II) and $\mathbf{m}_z < 0.5$ for state V, which increase with increasing magnetic field. 

Two spin eigenmodes are excited for the external microwave magnetic field along in-plane direction, which can be found in the spin excitation spectrum of skyrmion (Fig.~\ref{fig3} (a)). The lower frequency mode is the CCW mode, and the higher frequency mode corresponds to the CW mode. 
\begin{figure*}[!htb]
	\begin{center}
		\epsfig{file=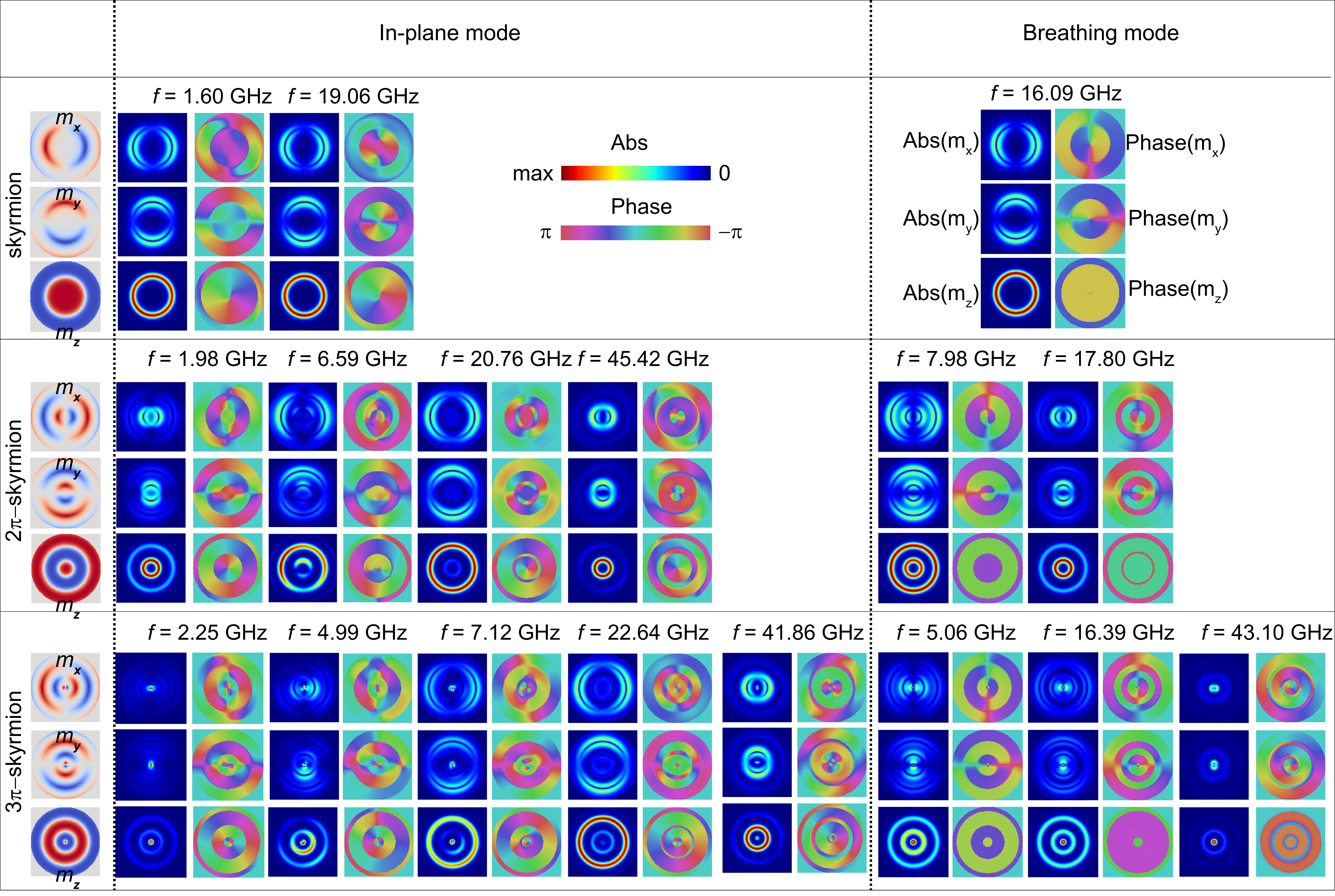,width=16 cm} \caption{The static magnetization component $\mathbf{m}_x$, $\mathbf{m}_y$ and $\mathbf{m}_z$ (left column) and the profile of the FFT power and phase of magnetization components in different frequency for skyrmion, $2\pi$ skyrmion and $3\pi$ skyrmion, including in-plane mode (middle column) and breathing mode (right column).}
		\label{fig4}
	\end{center}
\end{figure*}
We first consider the skyrmion excitation modes without applying the external magnetic field. The spatial profiles of the amplitude $Abs(m_i)$ ($i=x, y, z$) and phase $Phase(m_i)$ for dynamic magnetization components of skyrmion are plotted in the first row and middle column of Fig.~\ref{fig4}. It is characterized by the out-of-plane magnetization component $\mathbf{m}_z$ localized around the ring of skyrmion composed by in-plane magnetization. The amplitude of CCW or CW modes form a ring around the skyrmion edge, and the phase changes continuously from -$\pi$ to $\pi$, which means the dynamic magnetization oscillations are not in phase in the nanodot. With increasing magnetic field until 280 mT, frequencies of two azimuthal modes increases and then decreases due to the interaction between the skyrmion and nanodot boundary. We will discuss this phenomenon later. When the magnetic field is larger than 280 mT, the skyrmion state collapses to a uniform state (IV), the spin wave excitations of this single domain sates are not indicated in Fig.~\ref{fig3} (a). For $2\pi$ skyrmion, there are four spin wave excitation modes, two low frequency modes and two high frequency modes. The spatial profiles of $Abs(m_i)$ and $Phase(m_i)$ for dynamic magnetization at 0 mT are indicated in the second row and middle column in Fig.~\ref{fig4}, which shows that in low frequency $f=1.98$ GHz, the amplitude characterized by $\mathbf{m}_z$ formed a bright ring in the center and a slightly bright ring at the outer part in the nanodisk. Two rings formed in the nanodot at $f=6.59$ GHz and only one formed at the outer part ($f=20.76$ GHz) or at the inner part ($f=45.42$ GHz) in the nanodot. The dynamic magnetization oscillations are not in phase, where the phase changes continuously in each magnetization parts separated by domain wall in $2\pi$ skyrmion (Fig.~\ref{fig1} (b)). The frequencies of lower frequency mode decrease as a function of magnetic field, while the frequencies of higher frequency mode increase. The frequency is not shown when $2\pi$ skyrmion transforms to an another skyrmion state with $S_n=-1$ (VI) for that the frequency of spin wave excitation mode exceeds the cut-off frequency of excitation field. Compared to skyrmion and $2\pi$ skyrmion, the in-plane excitation modes for $3\pi$ skyrmion are more complicated, where five spin wave excitation modes exist. The spatial profiles of $Abs(m_i)$ and $Phase(m_i)$ for dynamic magnetization at 0 mT are indicated in the third row and middle column. Depending on the frequency excited, different rings are formed characterized by $\mathbf{m}_z$. In addition, the profiles of $Phase(m_i)$ for $3\pi$ skyrmion are still not in phase and the phase changes continuously in each magnetization part separated by domain walls in $3\pi$ skyrmion (Fig.~\ref{fig1} (c)). When magnetic field exceeds 60 mT, the spin excitation modes are similar to that of $2\pi$ skyrmion, then the excitation modes of skyrmion (VI) appear with $B_{ext}>280$ mT.

\begin{figure*}[!htb]
	\begin{center}
		\epsfig{file=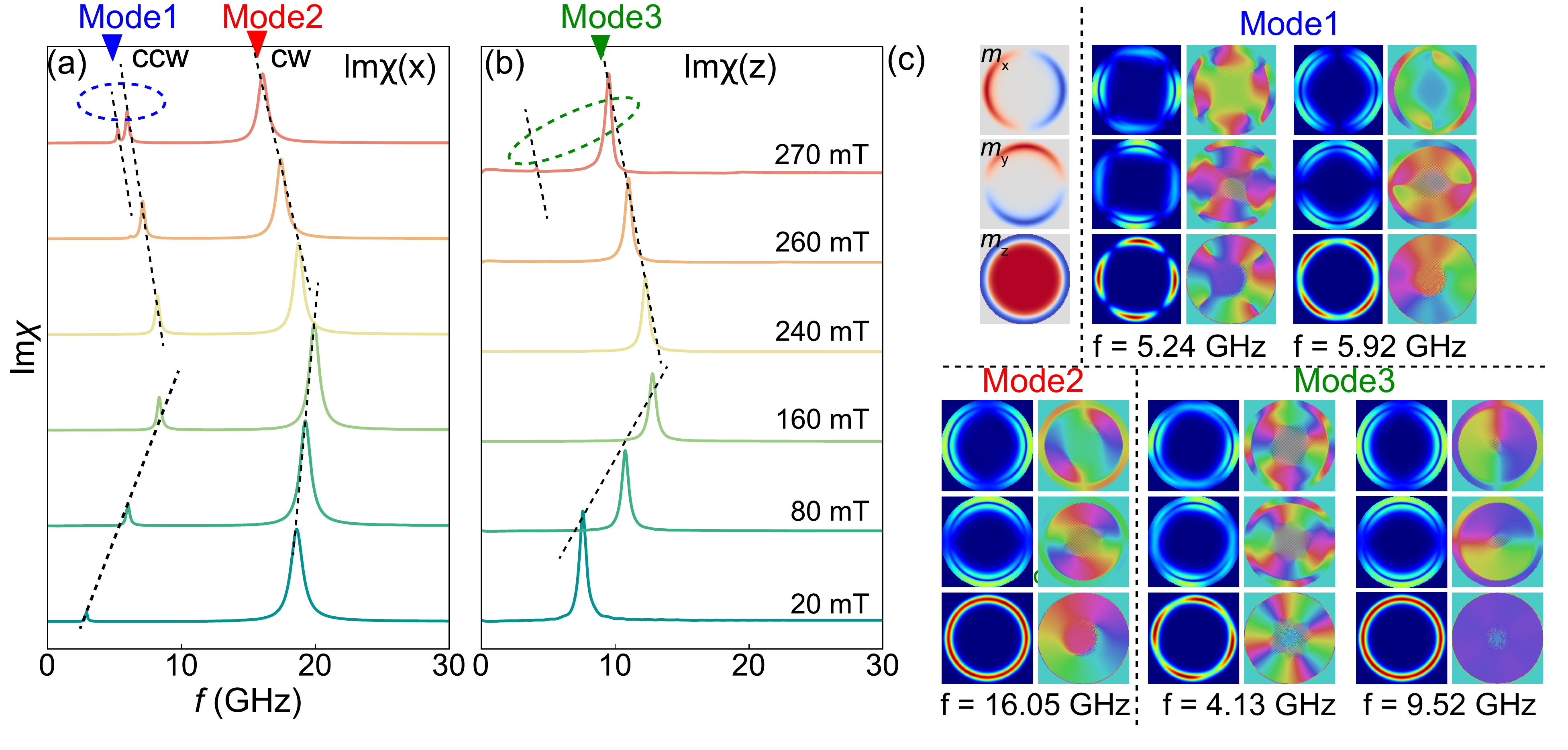,width=16 cm} \caption{Imaginary parts of the dynamical magnetic susceptibilities (a) lm$\chi$(x) and (b) lm$\chi$(z) as a function of frequency for several perpendicular magnetic field for a skyrmion confined in the nanodot, the activation magnetic field is applied along $x$ and $z$ direction, respectively. These modes are marked as Mode1 (CCW mode), Mode2 (CW mode) and Mode3 (Breathing mode). (c) Spatial profiles of amplitude and phase distribution for the three excitation modes with $B_{ext}=270$ mT.}
		\label{fig5}
	\end{center}
\end{figure*}
In Fig.~\ref{fig3} (b), a particular mode can be found in the excitation spectrum for skyrmion, which corresponds to breathing mode with skyrmion expansion and contraction periodically. Without considering the external magnetic field, the profile of amplitude and phase for dynamic magnetization component are depicted in the first row and right column of Fig.~\ref{fig4}. It shows that the out-of-plane magnetization components is localized in the domain wall of skyrmion which formed a ring, and the phase of breathing is in phase. With increasing magnetic field, the frequency increases and then it decreases, whereas the skyrmion radius increases. In a critical field, the skyrmion edge is localized near the nanodot edge and related edge effects appears. We will talk the edge effects later. For $2\pi$ skyrmion, there are two main modes in the excitation spectrum which are all corresponding to breathing mode. Whereas the area of the mode localization are different at 0 mT (second row and right column in Fig~\ref{fig4}), as well as the phase distribution. The amplitude of low frequency mode is localized at the two domain walls of $2\pi$ in the nanodot, while the high frequency mode is mainly localized in the inner domain of $2\pi$ skyrmion. The oscillation mode for low frequency is not in phase and it is in phase for high frequency mode. Applied magnetic field induces the increasing of excitation frequency for high frequency mode and decreasing for low frequency mode. At critical field, only one breathing mode depicted for state VI. Three breathing modes are excited for $3\pi$ skyrmion, the excitations are localized in three domain walls region of $3\pi$ skyrmion at $f=5.06$ GHz and the phase distribution shows that the oscillations of magnetization are not in phase. When $f=16.39$ GHz, the spatial profile of amplitude is similar to at $f=5.06$ GHz, while the spatial profiles of phase are quite different. The phase distribution at $f=5.06$ GHz shows that the inner part and the outer part separated by domain walls are in phase, while the middle part is opposite. Whereas the phase of inner part is opposite to that of the middle part and the outer part at $f=16.39$ GHz. For a high frequency breathing mode at $f=43.10$ GHz, the spatial profile of amplitude is mainly localized in the inner part, and the spatial profiles of the phase are same. It is noted that the numbers of modes increases with a larger $k$, and the phase distributions exhibit circular symmetry whether the breathing is in phase or not.

In Fig.~\ref{fig5}, we present a detailed microwave absorption spectra (a) lm$\chi$(x) and (b) lm$\chi$(z) of skyrmion when the excitation field is applied along $x$ and $z$ direction, respectively, as a function of microwave frequency for different perpendicular magnetic field. Two in-plane modes are marked as Mode1 and Mode2 corresponding to CCW mode and CW mode, and the breathing mode is marked as Mode3. Note that the excitation frequency of these three modes increase with increasing magnetic field until a critical field $B_{ext}=240$ mT. Then, they decrease as a function of field. The reason is that, increasing magnetic field will expand the skyrmion, when the edge of skyrmion is localized near the nanodot edge and related edge effects appears. In a particular magnetic field, lm$\chi$(x) and lm$\chi$(z) split a low frequency mode when $B_{ext}=270$ mT. The corresponding spatial profile of amplitude and phase of dynamic magnetization are shown in Fig.~\ref{fig5} (c), which shows that the skyrmion edge almost reaches to the boundary of nanodot. The left column of Fig.~\ref{fig5} (c) shows the spatial distribution of magnetization components $\mathbf{m}_x$, $\mathbf{m}_y$ and $\mathbf{m}_z$. For Mode1, which is related to two frequency $f=5.24$ GHz and $f=5.92$ GHz, the amplitude characterized by $\mathbf{m}_z$ formed a ring with four light parts and four dark parts. This is quite different from the spatial distribution of amplitude at 0 mT where a ring formed (Fig~\ref{fig4}). In addition, the phase for Mode1 changes continuously from -$\pi$ to $\pi$ for several times. A low frequency breathing modes appears for Mode3, the breathing at split frequency $f=4.13$ GHz are not in phase compared to the breathing at $f=9.52$ GHz, the spatial distribution of amplitude is not formed a uniform ring as well.

\begin{figure}[!htb]
	\begin{center}
		\epsfig{file=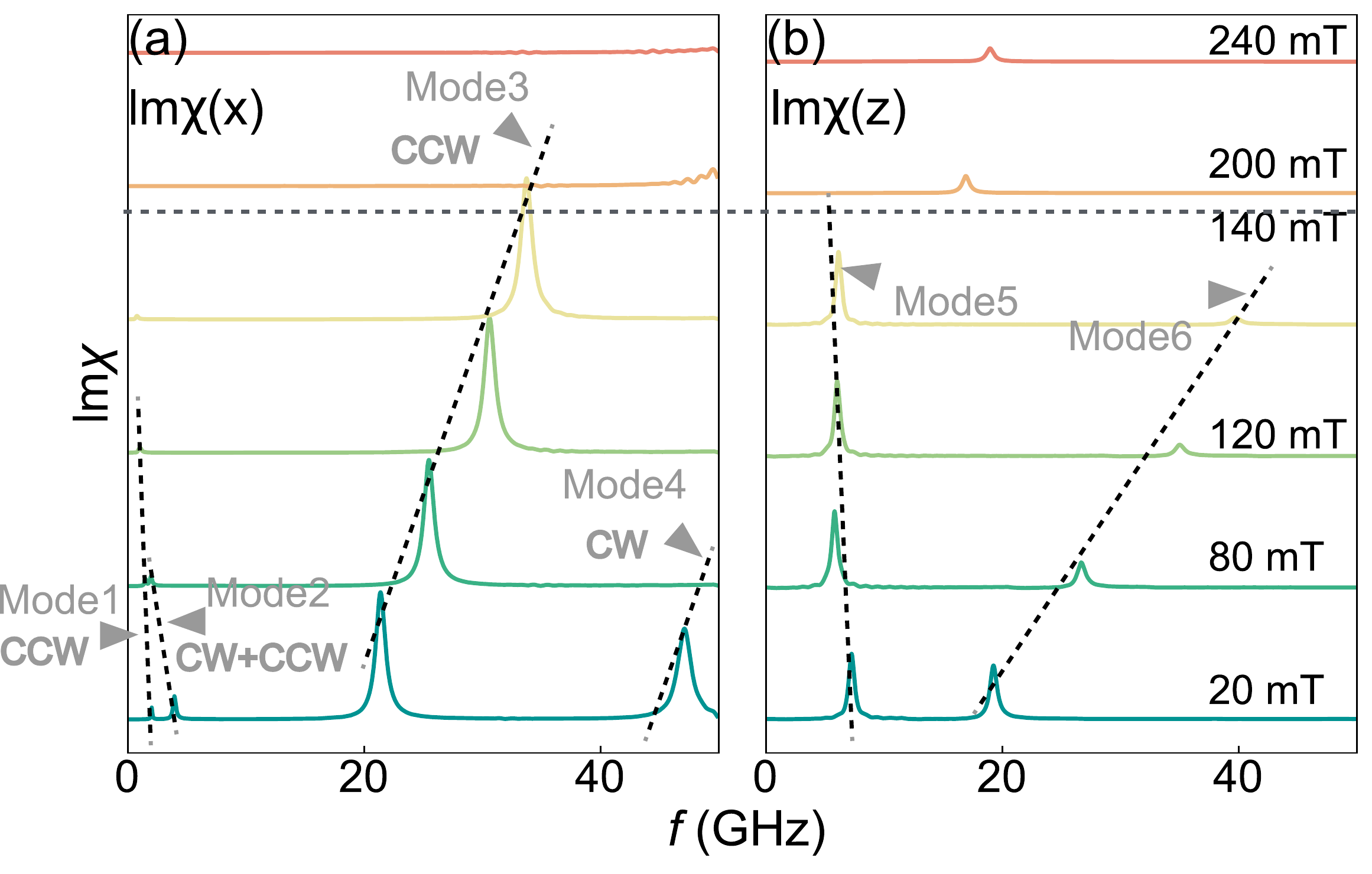,width=8 cm} \caption{Imaginary parts of the dynamical magnetic susceptibilities (a) lm$\chi$(x) and (b) lm$\chi$(z) for $2\pi$ skyrmion as a function of frequency, several magnetic fields are applying. Four in-plane modes are defined (Mode1, Mode2, Mode3 and Mode4) and two breathing modes are defined (Mode5, Mode6). The horizontal dotted line separates $2\pi$ skyrmion state (II) and state VI.}
		\label{fig6}
	\end{center}
\end{figure}
\begin{figure}[!htb]
	\begin{center}
		\epsfig{file=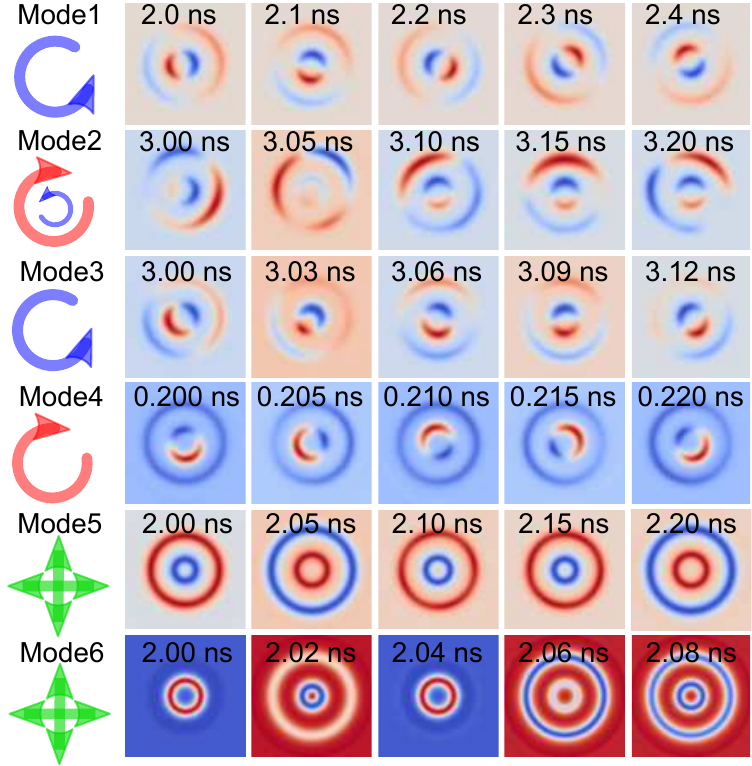,width=8 cm} \caption{Snapshots of the six excitation modes (Mode1-6) at different times marked in Fig.~\ref{fig6} of $2\pi$ skyrmion, which are activated by in-plane microwave magnetic field (Mode1-4) and out-of-plane magnetic field (Mode5-6) under an external perpendicular magnetic field $B_{ext}=20$ mT. Blue, red and green arrows represent CCW, CW and breathing modes, respectively.}
		\label{fig7}
	\end{center}
\end{figure}
For $2\pi$ skyrmion, the imaginary parts of the dynamical magnetic susceptibilities (a) lm$\chi$(x) and (b) lm$\chi$(z) for $2\pi$ skyrmion, as a function of frequency at several magnetic fields, are depicted in Fig.~\ref{fig6} . Four in-plane modes marked as Mode1, Mode2, Mode3 and Mode4 are depicted in Fig.~\ref{fig6} (a) from low frequency to high frequency. With increasing the magnetic field, the frequencies of two low frequency excitation modes (Mode1 and Mode2) decrease, while the frequency of two high frequency excitation modes increases. Two breathing modes are marked as Mode5 and Mode6, which depicted in Fig.~\ref{fig6} (b). Mode5 decreases slightly as a function of external magnetic field, whereas the Mode6 increases sharply with external magnetic field. When the $2\pi$ skyrmion collapses to a skyrmion (VI) at 180 mT, two azimuthal modes are not shown and only one breathing mode is depicted, and the excitation frequency increases with increasing magnetic field. In Fig.~\ref{fig7}, the snapshots of these six modes are shown for in-plane microwave magnetic field (Mode1-4) and out-of-plane microwave field (Mode5-6) when $B_{ext}=20$ mT. They are activated via application of a microwave magnetic field $B_i(t)=B_i\sin(2\pi ft)$ ($i=x, z$) with a corresponding excitation frequency as the time-dependent magnetic field $\mathbf{B}(t)$. For the purpose of visualization, we depicted the profiles of net magnetization in $z$ direction $\delta \mathbf{m}_z(t)$ defined as
\begin{equation}
\mathbf{m}_z(t)=\mathbf{m}_z(0)+\delta\mathbf{m}_z(t).
\end{equation}
We use the previously relaxed $2\pi$ skyrmion state as the equilibrium state $\mathbf{m}_z(0)$. The results show that Mode1 and Mode3 are CCW mode, and the rotation of inner and outer rings of $2\pi$ skyrmion are in opposite phase. For Mode2, the inner ring rotates with CCW mode, while the outer ring rotates with CW mode. Whereas only the inner ring is CW mode for Mode4. Two breathing modes (Mode5,6) depcit that the $2\pi$ skyrmion expands and shrinks periodically, where the inner skyrmion and outer skyrmion are breathing not in phase for Mode5 and in phase for Mode6.

\begin{figure}[!htb]
	\begin{center}
		\epsfig{file=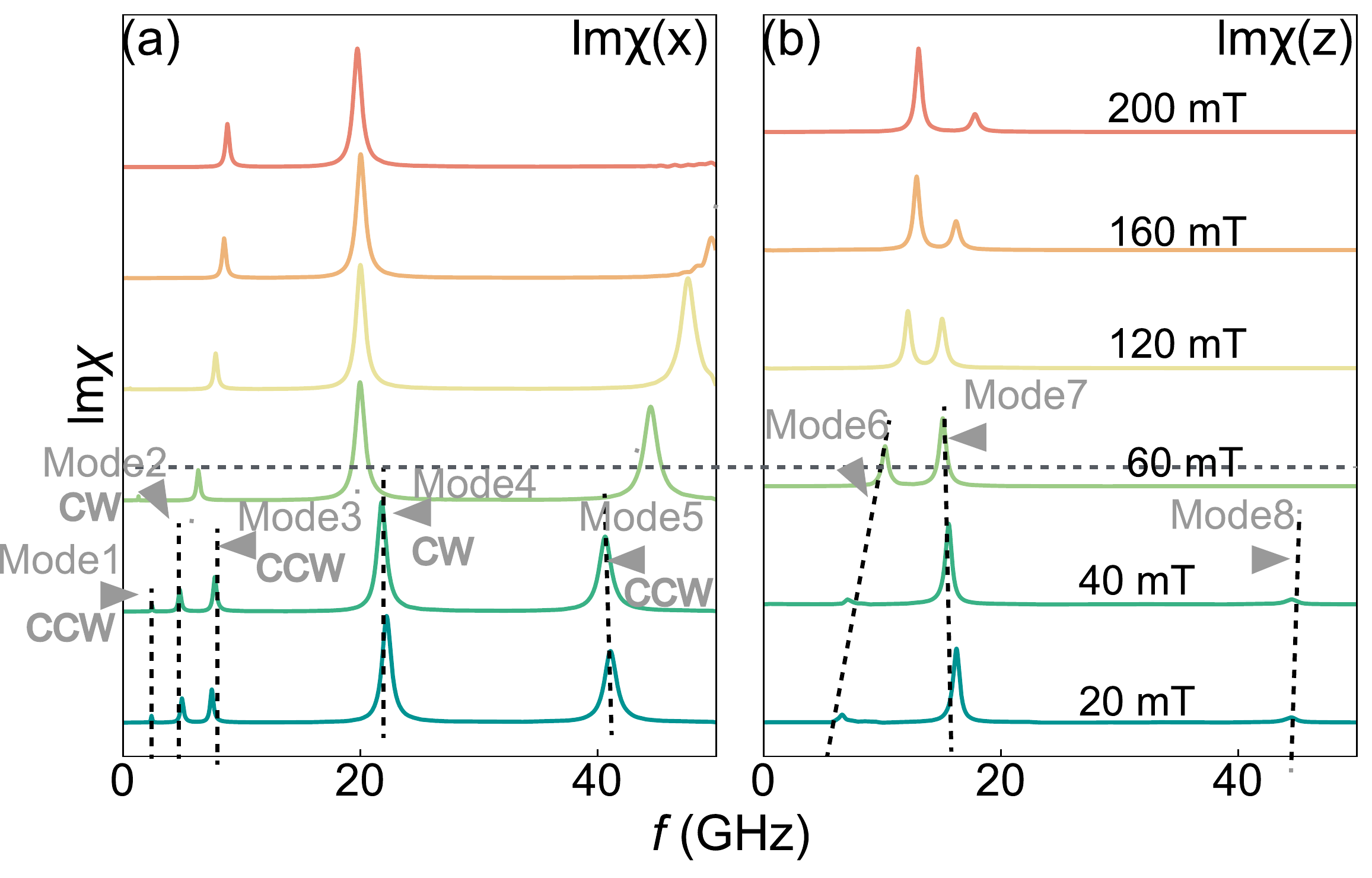,width=8 cm} \caption{Imaginary parts of the dynamical magnetic susceptibilities (a) lm$\chi$(x) and (b) lm$\chi$(z) for $3\pi$ skyrmion as a function of frequency, several magnetic fields are applying. Five in-plane modes are defined (Mode1, Mode2, Mode3, Mode4 and Mode5) and three breathing modes are defined (Mode6, Mode7, Mode8). The horizontal dotted line separates $3\pi$ skyrmion state (III) and state V.}
		\label{fig8}
	\end{center}
\end{figure}
\begin{figure}[!htb]
	\begin{center}
		\epsfig{file=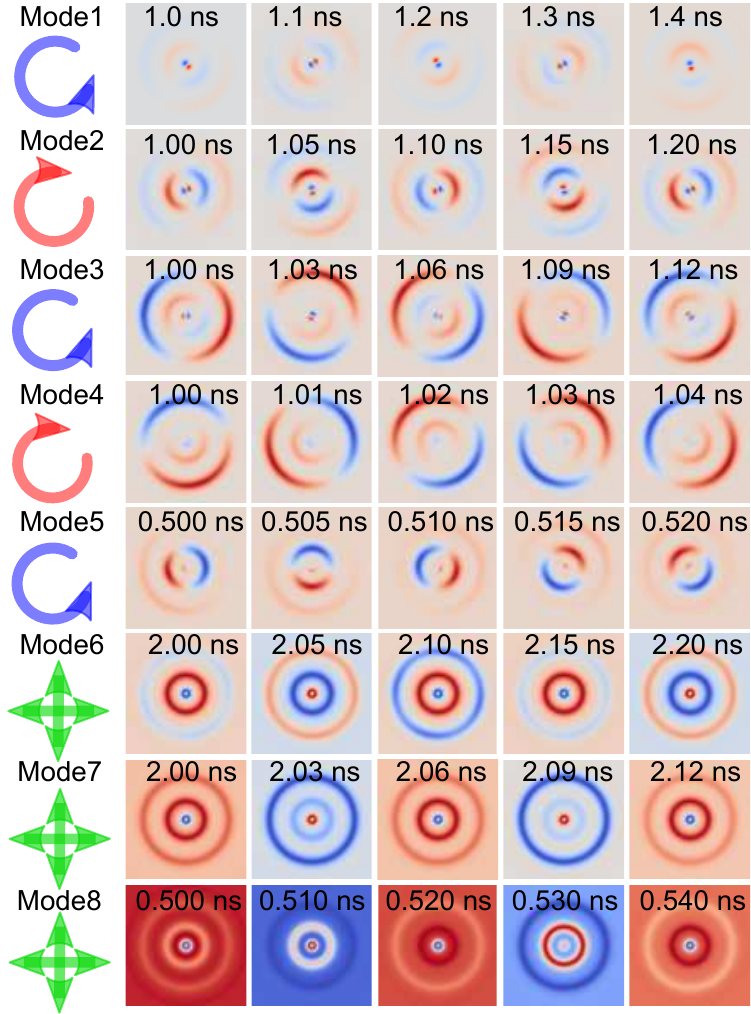,width=8 cm} \caption{Snapshots of the eight excitation modes (Mode1-8) at different times marked in Fig.~\ref{fig8} of $3\pi$ skyrmion, which are activated by in-plane microwave magnetic field (Mode1-5) and out-of-plane magnetic field (Mode6-8) under an external perpendicular magnetic field $B_{ext}=20$ mT. Blue, red and green arrows represent CCW, CW and breathing modes, respectively.}
		\label{fig9}
	\end{center}
\end{figure}
Compared skyrmion or $2\pi$ skyrmion, the spin wave excitation modes are more richer for $3\pi$ skyrmion. The imaginary parts of the dynamical magnetic susceptibilities (a) lm$\chi$(x) and (b) lm$\chi$(z) for $3\pi$ skyrmion are shown in Fig.~\ref{fig8}, as a function of frequency in different magnetic field. When the applied external magnetic field is under 70 mT, the $3\pi$ skyrmion are stabilized. From low frequency to high frequency, five in-plane excitation modes are defined (Mode1-5) and three breathing modes are depicted (Mode6-8). We find that the frequency of Mode1 and Mode2 decreases slightly with magnetic field, while it increases slightly for Mode3. For Mode4 and Mode5, the frequency decreases as a function of field. Fig.~\ref{fig8} (b) depicts that the frequency of Mode 6 and Mode8 increases opposite to the decrease of Mode7 with increasing magnetic field. Increasing magnetic field induces the collapse of $3\pi$ skyrmion to $2\pi$ skyrmion (V), where the excitation modes are similar to the results shown in Fig.~\ref{fig6}. It is worth noted that state V and state II are all composed of two skyrmion nested together, where only the polarities are opposite. Thus, we only focus the excitation modes for $3\pi$ skyrmion here. The corresponding snapshots for these eight excitation modes are shown in Fig.~\ref{fig9}, for in-plane microwave magnetic field (Mode1-5) and for out-of-plane microwave magnetic field (Mode6-8) when $B_{ext}=20$ mT. Under and in-plane microwave with eignfrequency of Mode1-5, the CCW mode and CW mode appear alternatively. Mode6-8 are all corresponding to breathing modes under perpendicular microwave magnetic field. Depending on the excitation area in $3\pi$ skyrmion, the net magnetization of $\mathbf{m}_z$ are lighted in different rings. For Mode6, the breathing phases of inner and outer rings are opposite to the middle ring, and for Mode7, the innder ring breathes in opposite phase compared to the middle and outer rings. While for Mode8, three rings are in same phase. These results show that the spin wave excitations depend on the magnetization configurations and excitation area.

\section{Summary}
In summary, we have determined spin excitation spectrum of $k\pi$ skyrmion in nanodot under perpendicular magnetic field, and shown the field induced transformations of different skyrmion-like topological magnetization structures. We investigated the transition states and related size change as a function of magnetic field. Moreover, we have found that with a larger $k$, the number of spin excitation modes increases, either for in-plane rotation modes or for the out-of-plane breathing modes, as the number of the ring $k$ increases with $k\pi$ skyrmion. Under an external magnetic along $z$-direction, the magnetization for different parts of $k\pi$ skyrmion expands or reduces depending on the magnetization orientation. Field induced magnetization profile change modulates the frequency of excitation modes for $k\pi$ skyrmion, which decreases or increases as a function of magnetic field. Therefore, using spin wave excitation modes of $k\pi$ skyrmion, the theoretical and experimental investigation and classification of these skyrmion-like structures are possible. These finding may open a promising application in spintronic devices or using in the magnonics. 

\section*{Acknodledgement}
This work is supported by National Science Fund of China (Grants No. 11574121 and No. 51771086)

\bibliography{reference}

\end{document}